# Materials selection rules for amorphous complexion formation in binary metallic alloys

**Jennifer D. Schuler [a], Timothy J. Rupert [a,b]**

[a] Department of Chemical Engineering and Materials Science, University of California, Irvine, CA, 92697, USA
[b] Department of Mechanical and Aerospace Engineering, University of California, Irvine, CA, 92697, USA

**Abstract**

Complexions are phase-like interfacial features that can influence a wide variety of properties, but the ability to predict which material systems can sustain these features remains limited. Amorphous complexions are of particular interest due to their ability to enhance diffusion and damage tolerance mechanisms, as a result of the excess free volume present in these structures. In this paper, we propose a set of materials selection rules aimed at predicting the formation of amorphous complexions, with an emphasis on (1) encouraging the segregation of dopants to the interfaces and (2) lowering the formation energy for a glassy structure. To validate these predictions, binary Cu-rich metallic alloys encompassing a range of thermodynamic parameter values were created using sputter deposition and subsequently heat treated to allow for segregation and transformation of the boundary structure. All of the alloys studied here experienced dopant segregation to the grain boundary, but exhibited different interfacial structures. Cu-Zr and Cu-Hf formed nanoscale amorphous intergranular complexions while Cu-Nb and Cu-Mo retained crystalline order at their grain boundaries, which can mainly be attributed to differences in the enthalpy of mixing. Finally, using our newly formed materials selection rules, we extend our scope to a Ni-based alloy to further validate our hypothesis, as well as make predictions for a wide variety of transition metal alloys.

# 1. Introduction

Internal interfaces can significantly influence material behaviors such as plastic deformation [1], fracture [2], and corrosion [3], and engineering these interfaces can in turn lead to improved performance [4, 5]. An exciting new concept for the design and control of interfacial properties are "complexions," interfacial structures that are in thermodynamic equilibrium and have a stable, finite thickness [5-7]. Complexions can be considered quasi-2D "phases" that only exist at an interface, surface, or grain boundary [8]. Similar to bulk phases, complexions can be described with thermodynamic parameters and can even undergo phase-like transitions in response to alterations of external variables such as temperature, pressure, chemistry, and grain boundary character [8]. Since their existence is dependent on the neighboring crystalline grains, complexions do not technically adhere to the Gibbs definition of a phase and thus are considered with a separate terminology [8, 9].

Dillon et al. [6] developed a convention to classify complexions into six different types according to thickness, structural ordering, and composition. The six types suggested were: (I) sub-monolayer segregation, (II) clean, undoped grain boundaries, (III) bilayer segregation, (IV) multilayer segregation, (V) nanoscale intergranular films, and (VI) wetting films. This continuum of complexion types can be subdivided into ordered or disordered. Complexion types I-IV have crystalline structure and are classified as ordered, whereas types V and VI can assume either an ordered or disordered structure. The disordered versions of type V or VI complexions can be classified as amorphous intergranular films (AIFs) [6]. Different complexion types have been shown to dramatically influence material behavior and have been deemed the root cause behind several previously unexplained phenomena. Ordered bilayer complexions were found to explain liquid metal embrittlement in Cu-Bi [10] and Ni-Bi [11] due to the segregation of Bi to the grain boundaries which stretches the intergranular atomic bonds to near the breaking point, making them very fragile. Similarly, Ga segregates to the grain boundaries in Al-Ga to form an ordered trilayer complexion that also has an embrittling effect [12]. Conversely, AIFs have been shown in some situations to improve damage tolerance due to the excess free volume present in the amorphous grain boundary structure [13-15]. Atomistic simulations have shown that the amorphous-crystalline interfaces that bound AIFs can attract dislocations [15]. AIFs can also act as a toughening feature to delay intergranular crack formation and propagation [13, 14], as well as increase radiation tolerance by acting as an efficient and unbiased sink for point defects [16]. Experimental studies support these findings, with nanocrystalline Cu-Zr containing AIFs demonstrating enhanced strength and ductility compared to the same alloy with ordered grain boundaries [17, 18]. Amorphous complexions have also been shown to dramatically increase diffusion, which can cause abnormal grain growth [6] and solid-state activated sintering [19]. Solid-state activated sintering, referring to improved densification rates that occur below the solidus temperature, has been observed in both metallic and ceramic systems. The addition of a small amount of sintering aid element creates disordered intergranular films that act as a pathway for improved diffusion below the bulk eutectic temperature [13-16, 19, 20]. In addition, AIFs were recently found to stabilize nanocrystalline grain structures against grain growth at elevated temperatures, with a nanocrystalline Cu-Zr alloy remaining nanostructured even after a week at 98% of its melting temperature [18].

Due to the enhanced performance imparted by AIFs, the application of these unique grain boundary structures to a wider array of alloys would be advantageous. The hypothesis of surface premelting promoted interest in stable interfacial films [21], which lead to thermodynamic descriptions of 2D-interfacial films that undergo phase transformations [22-24]. Complexions have since been extensively studied in ceramics [6, 8, 25] and multicomponent metallic systems where AIFs are accessible [26, 27]. Advancement of the thermodynamic theories behind complexions has even allowed for the development of grain boundary phase diagrams that connect structural transitions at an interface with alloy composition and temperature, emphasizing their phase-like behavior [7, 19].

While the theoretical framework behind AIF formation is well-developed, the implementation of this concept to new alloy systems has been limited. The experimental study and application of these features has been largely relegated to ceramics where AIFs have been extensively observed [6, 28], or in alloys where AIFs were already suspected, such as those alloys that exhibit the AIF-driven behavior of solid-state activated sintering [29]. Development of a general set of materials selection rules using readily available material parameters to predict material systems in which AIFs are possible would be powerful. The history of amorphous materials research can serve as an instructive example of this concept. In 1932, Zachariasen [30] offered a critical discussion of the structure of glassy ceramics and suggested general guidelines for materials selection, prompting a flurry of discoveries and advancements built upon these guidelines. Despite their rudimentary nature, Zachariasen's rules are recognized as one of the first attempts to systematically address glass forming ability, fundamentally influencing future research in the field [31]. Similarly, as interest began to build for amorphous metals, Inoue [32] suggested a set of three empirical rules that have since provided a preliminary guide for the development of new bulk metallic glasses (BMGs).

In this study, we propose materials selection rules for the promotion of AIFs that emphasize dopant segregation to grain boundaries and the creation of energetically favorable conditions for forming an amorphous region. To test the robustness of these rules, a variety of Cu-rich systems with contrasting thermodynamic parameters were selected and processed. Here, we focus on transition metal dopants in order to avoid complicating factors such as directional bonding, complex kinetics, and crystallographic anisotropy dependence that are characteristic of ceramic systems [19, 33, 34]. The behavioral patterns established by the inspection of the Cu-rich alloys are then extended to predict the complexion formation behavior of a new Ni-based alloy where AIFs have not yet been observed in prior work. In summary, the type of complexion formed at the grain boundaries of a polycrystalline binary metallic alloy can be controlled by an informed selection of enthalpy of segregation ($\Delta H^{seg}$), enthalpy of mixing ($\Delta H^{mix}$), and atomic radius mismatch, where AIF formation depends on dopant segregation to the grain boundary and the glass forming ability of the alloy.

## 2. Materials and Methods

The alloys used in this study were produced with magnetron co-sputtering using an Ar plasma in an Ulvac JSP 8000 metal deposition sputter tool. Sputtering was specifically chosen in order to create high purity samples. High-purity targets were obtained from Kurt Lesker with purities of 99.99 wt.% for Cu, 99.2 wt.% (inc. Hf) for Zr, 99.9 wt.% (exc. Zr) for Hf, 99.95 wt.% (exc. Ta) for Nb, 99.95 wt.% for Mo, and 99.99 wt.% for Ni. In addition, deposition was only performed after a $10^{-7}$ mtorr base chamber pressure was achieved to further minimize impurity incorporation into the films. The films were deposited at 400 °C using an Ar pressure of $1.5 \times 10^{-3}$ mtorr with sample stage rotation during deposition in order to achieve a uniform film. The metals were co-deposited onto Cu or Ni substrates which had been polished to a mirror surface finish prior to deposition. Films were deposited onto sheets of the primary alloying element in order to eliminate unwanted chemical reactions between the thin film and substrate during subsequent thermal processing. A summary of the key deposition parameters, processing details, and film information are presented in Table 1.

Since the average thickness of a transmission electron microscope (TEM) sample must generally be <100 nm in order to achieve electron transparency [35], very small grains can overlap, introducing uncertainty to structural and chemical analysis [36]. In order to minimize this issue, high sputtering temperatures were chosen in order to increase atom mobility and maximize grain size at deposition, as well as suppress the growth of a void-filled film [37]. A micrometer-scale film thickness was also targeted since the maximum grain size achievable in a thin film is typically tied to the film thickness [38]. After deposition, all samples were annealed under vacuum at 500 °C for 24 hr to promote further grain growth and allow for segregation of dopants to the grain boundary to achieve chemical equilibrium.

Target alloy compositions were chosen far from any intermetallic compositions but above the solid solubility limits in order to minimize the unwanted precipitation of second phases while still promoting grain boundary segregation. Different complexion types can be accessed by modulating processing conditions, such as temperature and pressure, to control complexion type transformations [8], with higher temperatures promoting the formation of thicker AIFs [7]. In order to maximize AIF formation, the samples were heated to ~0.92$T_{solidus}$ of the alloy at the measured composition (900 °C for Cu-Zr, 915 °C for Cu-Hf, and 1000 °C for both Cu-Nb and Cu-Mo), held for 1 minute and then rapidly quenched to preserve any thermodynamically stable interfacial structures that are only achievable in the heated state. In order to execute the heating and quenching steps without oxidation, the samples were sealed under vacuum in high purity quartz tubes, suspended in a vertically-oriented tube furnace for the high temperature annealing, and then dropped into a water bath in under 1 s for quenching. The 500 °C anneal for 24 hr permits long range diffusion of the dopants. After this, the diffusion length scales calculated for the ~0.92$T_{solidus}$ anneal for 1 minute are on the scale of hundreds of nanometers for each alloy, providing ample opportunity for dopants already localized at the grain boundary post the 500 °C anneal for 24 hr to reorder across the nanometer scale. This local reorganization thus permits dopants segregated to the grain boundary to reorder into a thermodynamically favorable state, such as an AIF.

TEM samples were created using the focused ion beam (FIB) lift-out technique on an FEI dual beam Quanta 3D microscope using $Ga^+$ ions. To reduce ion beam damage, all TEM samples received a final polish with a low power 5 kV beam to remove surface amorphization and minimize damage caused by the beam. Bright field (BF) TEM images and selected area electron diffraction (SAED) patterns were collected using a Philips CM-20 operating at 200 kV. The average grain sizes of the alloys were determined by measuring the areas of at least 100 grains and calculating the average equivalent circular diameter. High resolution TEM (HRTEM) was performed on an FEI Titan at 300 kV. Energy-dispersive X-ray spectroscopy (EDS) and high angle annular dark field (HAADF) scanning TEM (STEM) were collected on the same microscope at 300 kV. Fresnel fringe imaging was used to identify interfacial films as well as to ensure edge-on orientation of the grain boundary during imaging [39].

## 3. Results and Discussion

### 3.1. Proposed Materials Selection Rules for AIF Formation

We hypothesize that two key requirements must be satisfied for a nanoscale amorphous complexion to form in a binary metallic alloy. First, sufficient excess dopant needs to be present at the grain boundary in order to drive AIF formation. In situ TEM heating experiments have shown that grain boundary premelting is vanishingly difficult in pure monotonic metals, with an ordered boundary structure persisting to at least 99.9% of the melting temperature [40]. Alternatively, the addition of a segregating dopant can make grain boundary premelting conditions accessible at much lower temperatures (e.g., 60-85% of the melting temperature for W-rich alloys [29]), explaining solid-state activated sintering [41]. The enthalpy of segregation, $\Delta H^{seg}$, describes whether it is energetically favorable for a dopant element to segregate to the grain boundary in a polycrystalline system, with a positive value denoting a propensity for segregation and a negative value denoting a preference for depletion of the dopant at the grain boundary [42].

Murdoch and Schuh [42] developed a catalogue of $\Delta H^{seg}$ values using a Miedema-type model for a large number of binary alloy combinations in order to further understand the role of this parameter in stable nanocrystalline alloy design. By lowering the grain boundary energy through dopant segregation, the thermodynamic driving force for grain growth is mitigated, allowing these materials to retain their desirable nanocrystalline structure even when exposed to elevated temperatures [43]. The theoretical framework to predict stable nanocrystalline materials using a thermodynamic stabilization route has made considerable progress in recent years [42, 44-50]. Darling et al. [51] also contributed to this field by calculating stability maps for the solute composition needed to minimize the excess grain boundary energy for a given grain size and temperature. Both types of studies provide a firm foundation for elemental selections when designing thermally-stable nanocrystalline alloys by utilizing grain boundary segregation. Similarly, the first requirement for nanoscale AIF formation is that $\Delta H^{seg}$ must be positive to ensure sufficient dopant is situated at the grain boundary.

The second requirement for AIF formation is that it must be energetically favorable for the grain boundary to assume an amorphous structure with a stable thickness and chemical composition. Nanoscale AIF formation is energetically favorable when the free energy penalty associated with the formation of a disordered film of a certain thickness is less than the reduction in interfacial energy caused by the replacement of the original crystalline grain boundary with two new amorphous-crystalline interfaces, as summarized in Equation 1 [19]:

$$\Delta G_{amorph} \cdot h < \gamma_{GB} - 2\gamma_{cl} \equiv \Delta\gamma \qquad (1)$$

$\Delta G_{amorph}$ refers to the volumetric free energy penalty for an undercooled amorphous film at a given alloy composition, $h$ is the film thickness, $\gamma_{gb}$ is the excess free energy of the original crystalline grain boundary, and $\gamma_{cl}$ is the excess free energy of the crystalline-amorphous interface. From Equation 1, it is advantageous to have a small free energy penalty for the amorphous phase to promote AIF formation or alternatively sustain thicker AIFs. Due to their amorphous structure, AIFs bear a clear resemblance to BMGs. Prior work has even shown that the short-range structural order in the interior of an AIF is identical to a bulk amorphous phase [52]. As such, we propose that the materials selection rules used for the creation of BMGs can be instructive for nanoscale AIFs.

Three empirical guidelines, primarily introduced by Inoue [32], have been used to improve the glass forming ability (defined as the critical cooling rate needed to retain an amorphous structure during solidification from the melt) of materials for BMG production. First, multi-component alloys, usually consisting of three or more elements, increase the complexity and size of the possible crystalline structures, reducing the possibility of long range periodicity upon cooling [32, 53]. While ternary and higher alloys make the best BMGs, examples exist in binary systems as well, such as Cu-Zr [54]. Binary alloys were selected for this work in order to simplify the selection process and ensure segregation, since grain boundary enrichment is critical for nanoscale AIF formation in accordance with the first AIF selection rule. The prediction of segregation behavior in systems with multiple dopants is challenging, as the various dopants can compete for segregation sites and interact to influence the final microstructure [55]. In this study, we focus on binary systems in order to circumvent this complicating factor while still allowing for grain boundary enrichment.

Second, a large atomic radius mismatch between elements further hinders the formation of a crystalline structure by creating a high packing density in the amorphous structure, which impedes the free volume expansion necessary to form a crystalline structure [32, 53]. The atomic radius mismatch is defined as the difference in the metallic bonding radii [56] of the elements in the binary metallic alloy divided by the radius of the smaller element, where a value greater than 12% is preferential for BMG formation [53, 57], as provided in Equation 2:

$$\Delta r/r = (r_{larger} - r_{smaller})/r_{smaller} > 12\% \qquad (2)$$

Third, a negative $\Delta H^{mix}$ creates a thermodynamically favorable landscape that reduces the rate of crystal nucleation [32, 53]. A negative $\Delta H^{mix}$ refers to an exothermic solution where energy is released upon mixing, meaning bonding between differing elements is favorable. Conversely, a positive value refers to an endothermic solution where bonding between like elements is favorable [58]. A common empirical signature of a negative $\Delta H^{mix}$ is the presence of many intermetallic phases on the equilibrium phase diagram.

Thus, alloys with a both positive $\Delta H^{seg}$ as well as a combination of a negative $\Delta H^{mix}$ and atomic radius mismatch greater than 12% are promising candidates for AIF formation. Four Cu-rich, binary metallic alloys (Cu-Zr, Cu-Hf, Cu-Nb, and Cu-Mo) that exhibit dopant segregation and possess a range of $\Delta H^{mix}$ and atomic radius mismatch combinations were chosen in order to test these selection rules. Cu-Zr and Cu-Hf have a positive $\Delta H^{seg}$, as calculated using the Miedema method [42]. Cu-Nb is also expected to have a positive $\Delta H^{seg}$, due to previous modeling and experimental research that has shown Nb segregation and clustering at grain boundaries in Cu [59-61]. Experiments on Cu-Mo have shown that irradiation leads to Mo clustering at grain boundaries [62], also indicating a positive $\Delta H^{seg}$. In addition, Atwater and Darling [63] calculated a theoretical minimum grain boundary energy caused by Nb and Mo added to nanocrystalline Cu, further suggesting a thermodynamic propensity for dopant segregation to lower the grain boundary energy. In order to examine the second half of our materials selection requirements (promotion of an amorphous structure), alloys were chosen with different thermodynamic parameters. Cu-Zr and Cu-Hf have negative $\Delta H^{mix}$ values [64], whereas Cu-Nb and Cu-Mo have positive $\Delta H^{mix}$ values [51, 65]. Cu-Zr and Cu-Hf have several intermetallic phases with deep eutectics that can form whereas Cu-Nb and Cu-Mo do not [66], reflective of the $\Delta H^{mix}$ parameters in these systems. Additionally, all of the alloys except Cu-Mo have an atomic radius mismatch greater than 12%. Cu-Zr is a well-known glass former and was in fact the first binary BMG created [54], with evidence emerging that the high crystal-liquid interfacial free energy of this alloy is responsible for this behavior [67]. Cu-Hf has also exhibited reasonable glass forming ability [68].

As a result, we predict that Cu-Zr and Cu-Hf alloys can sustain nanoscale AIFs because dopant segregation is encouraged while favorable values of $\Delta H^{mix}$ and atomic radius mismatch promote the formation of an amorphous structure. Conversely, Cu-Nb and Cu-Mo are predicted to have ordered grain boundaries due to the low glass forming ability of these systems. Grain boundary segregation of the added dopants is expected in all four of the alloy systems. The key thermodynamic parameters and predictions are summarized in Table 2.

### 3.2. Characterization of Cu-rich Alloys

BF TEM images of the Cu-rich alloys after heat treatment are shown in Figure 1. All of the alloys exhibited equiaxed grains, with the average grain sizes and standard deviations presented in Table 1. Despite a deposition temperature of 400 °C, micrometer scale film thicknesses, a 500 °C anneal for 24 hr, and an annealing step at $0.92T_{solidus}$ (all of which should promote grain growth), the Cu-Zr, Cu-Hf, and Cu-Mo films were still nanocrystalline with average grain sizes of 99 nm, 47 nm, and 85 nm, respectively. This is in contrast to pure Cu films deposited using similar deposition conditions which exhibited substantial grain growth. Cu-Nb also exhibited some amount of grain boundary stabilization, although to a lesser degree with an average grain size of 468 nm in the ultrafine-grained regime. EDS elemental maps were collected to provide a preliminary understanding of the degree of grain boundary segregation and dopant distribution experienced by each alloy, as shown in Figure 2 and Figure 3. Figure 2 shows the accompanying HAADF STEM image to the EDS map for Cu-Zr, highlighting how the dopant concentration values are highest at the grain boundaries. Both Figures 2 and 3 show that dopant concentration is inhomogeneous and that segregation to the grain boundaries occurs, in agreement with the positive $\Delta H^{seg}$ values for the four alloys.

While all of the alloys experienced dopant segregation, the grain boundary structures differed significantly. TEM inspection of interfaces in the Cu-Zr and Cu-Hf system are presented in Figure 4. The SAED insets in Figures 1(a) and (b) confirm that the Cu-Zr and Cu-Hf films were solid solutions with no second phase precipitation, indicated by the presence of only the face centered cubic (FCC) Cu diffraction rings in the pattern [69, 70]. Figure 4(a) shows an HRTEM image of a ~2 nm thick AIF in the Cu-Zr alloy, with Figure 4(b) displaying the accompanying EDS line profile scan for that grain boundary. The Zr segregation is evident in the line profile, reaching a maximum value of 7 at.% Zr and dropping to approximately 1 at.% Zr in the grain interior. It is important to note that the interaction volume of the electron beam is likely larger than the grain boundary thickness, meaning the maximum Zr composition measured is an average of the AIF composition and the crystalline material next to it. The segregation observed here is similar to the behavior reported by Khalajhedayati and Rupert [18] in a Cu-Zr alloy with AIFs that was created through ball milling. Figures 4(c) and (d) present similar data for the Cu-Hf system, showing a 5 nm thick AIF that reaches a maximum dopant concentration of ~12 at.% Hf at the grain boundary but then the composition drops down as the line profile extends into the neighboring grains. Again, the overall trend of dopant segregation to the grain boundary is clear.

Also presented in Figures 4(a) and (c) are fast Fourier transform (FFT) patterns taken from HRTEM images of the grain boundary films and the neighboring grains. The FFTs of the adjoining grains show periodic spots around the center point, indicating the presence of crystalline order, which also appears in the HRTEM image as lattice fringes. In contrast, the FFTs of the grain boundary film are featureless, confirming the presence of an amorphous region. The thickness of the films in Figures 4(a) and (c) are constant along the grain boundary, suggesting that the films are in thermodynamic equilibrium with the two neighboring crystalline grains and can be classified as type V nanoscale AIFs. Work by Dillon and Harmer [25] on complexions in $Al_2O_3$ showed that the thickness along a wetting film can change significantly and tended to be much thicker (>10 nm in many cases) than the films found here, lending additional confidence to the classification of these films as a type V nanoscale AIFs and not type VI amorphous wetting films. It is important to note that the thickness of the observed AIFs varied from boundary-to-boundary and that some interfaces even appeared ordered without an amorphous complexion. This variety of thicknesses was also observed in ball milled Cu-Zr [17] and is likely due to variations in grain boundary character [71] as well as local fluctuations in Zr content.

Figures 5 and 6 show HRTEM images and EDS line profiles of representative grain boundaries in the Cu-Nb and Cu-Mo systems. Both systems had only atomically sharp grain boundaries with ordered structures. No AIFs were found even after the inspection of many boundaries. Figures 5(b) and 6(b) show EDS line scans of Cu-Nb and Cu-Mo grain boundaries with excess dopant being seen for each system, reaching 8 at.% Nb and 10 at.% Mo at the grain boundaries and dropping down inside the neighboring grain interiors. While dopant segregation was present at the grain boundaries, dopant-rich crystalline clusters were also found at both the grain boundaries and within the grain interiors for both alloys. For Cu-Nb, the clusters were typically ~30 nm in diameter. An HRTEM image of a Nb precipitate located at a grain boundary is shown in Figure 5(c), with the associated EDS line scan across the cluster in Figure 5(d), reaching a maximum value of 27 at.%. Nb. Cu-Mo formed smaller clusters that were ~5 nm in diameter. An HRTEM image of multiple Mo clusters is presented in Figure 6(c), with the associated EDS line scan across the cluster in Figure 6(d) showing a maximum composition of 17 at.% Mo. Again, it is likely that the compositions of the precipitates are higher due to the surrounding Cu being included in the beam interaction volume. The FFT insets in Figures 5(c) and 6(c) confirm the crystallinity of the clusters and neighboring grains. A summary of the grain boundary structures found in the Cu alloys is presented in Table 3.

The efficacy of dopant segregation in stabilizing grain size was particularly evident in the Cu-Zr, Cu-Hf, and Cu-Mo films, which remained nanocrystalline despite processing efforts to increase the grain size for easier TEM inspection. Such grain size stability, at temperatures as high as $0.92T_{solidus}$, has been documented for Cu-Zr [18] but is a new observation for the Cu-Hf and Cu-Mo systems. While stabilization through doping has been reported in systems such as Ni-W [73], Hf-Ti [74], and W-Ti [75], the annealing temperatures used were significantly lower than $0.90T_{melting}$ in these studies. Darling et al. [72] did report on a very stable nanocrystalline Fe-Zr alloy, with the stability attributed to Zr segregation. In the case of Cu-Zr and Cu-Hf shown here, it appears that stability at high temperatures is aided by AIF formation, since these features are the lowest energy structures available at such high temperatures and therefore fit into the thermodynamic theories of stabilization.

On the other hand, the Cu-Mo system is stabilized by a combination of grain boundary segregation as well as the presence of small precipitates, meaning both thermodynamic and kinetic stabilization are active. The kinetic contribution comes from Zener pinning caused by the dopant clusters [59, 76-78]. Clustering of Mo and the eventual precipitation of a second phase in a Cu-rich alloy has been previously reported due to the immiscibility of the added dopant [59, 60, 62, 79]. Similar behavior has been observed in Cu-Ta, an alloy system that also has a positive $\Delta H^{mix}$ and experiences dopant segregation [78, 80]. Finally the Cu-Nb alloy does not appear to be adequately stabilized, even though Nb segregates to the grain boundaries and precipitates do form. Kapoor et al. [59] also reported grain growth in Cu-Nb where the grain growth behavior was dependent on the Nb concentration, with lower percentages promoting grain growth. It is also possible that the larger size of the precipitates (tens of nm in diameter for Cu-Nb versus only a few nm in diameter for Cu-Mo) is responsible for the lack of stability, as a uniform distribution of many fine particles smaller than the critical precipitate radius is best for reducing grain boundary motion [81].

In summary, Cu-Zr and Cu-Hf both contained nanoscale AIFs after being quenched from a high annealing temperature, showing that a negative $\Delta H^{mix}$ and a large atomic radius mismatch promote such features. As hypothesized in our design rules, it is also clear that a two component alloy is sufficient for stabilizing AIFs. The difference between AIFs (requires two elements) and BMGs (usually have three or more elements) can perhaps be attributed to the different length scales over which an amorphous structure must be stable. AIFs only require disorder of a nanoscale region, while BMGs require disorder that extended over mm length scales. This suggests that compositions that can sustain AIFs should be more plentiful than those which can be used for BMGs.

Cu-Nb and Cu-Mo both have positive $\Delta H^{mix}$ values, but these two alloys are differentiated by one key materials selection metric: the atomic radius mismatch. Cu-Nb has an atomic radius mismatch of 14% while Cu-Mo has a mismatch of 8.6%. Cu-Mo therefore achieves neither of the criteria needed to sustain an amorphous film and only has ordered grain boundaries as expected. Despite Cu-Nb satisfying one of the empirical rules for BMG formation, this alloy only exhibited ordered grain boundaries structures. When only looking at the results from our Cu-rich alloys, it is impossible to confirm whether both a negative enthalpy of mixing and a large atomic size mismatch are needed, or whether the negative enthalpy of mixing criteria is enough to predict AIF formation with atomic size being a secondary consideration. However, a detailed discussion of the available literature in the next section can clarify this point.

### 3.3. Extension of Materials Selection Rules to New Alloys

To make a final determination of our materials selection rules, it is necessary to examine a larger collection of literature reports. Table 4 shows a summary of binary metallic alloys that have exhibited behaviors which can be attributed to complexion formation. The longstanding mystery of grain boundary embrittlement has recently been solved and attributed to ordered complexions in Ni-Bi [11], Cu-Bi [10], and Al-Ga [12]. These alloy systems have a positive $\Delta H^{mix}$ and therefore ordered complexions would be predicted, which agrees with experimental observations. Solid-state activated sintering is typically attributed to the presence of AIFs and has been observed for certain Mo-rich and W-rich alloys [29, 41]. Inspection of Table 4 shows that all of the alloys which experience activated sintering, and therefore likely contain AIFs, have negative $\Delta H^{mix}$ values, but some of these materials do not have large atomic size mismatches (one of the empirical rules to enhance glass forming ability). This observation shows that the BMG formation guidelines may be slightly different when applied to AIFs and used in conjunction with the other AIF formation requirements, including sufficient dopant segregation to the grain boundary and suppression of competing second phase nucleation, in that $\Delta H^{mix}$ is of primary importance while a negative a large atomic size mismatch may be a secondary consideration. In contrast, activated sintering was not observed for W-Cu [29], which agrees with our prediction that only ordered boundaries would be present due to the positive $\Delta H^{mix}$ of the system.

Inspection of the grain boundary structures in various Cu-rich alloys, as well as a critical review of literature data, allows us to finalize our materials selection rules. A positive $\Delta H^{seg}$ leads to dopant segregation while a negative $\Delta H^{mix}$ plus a large atomic size mismatch promotes AIF formation, where a large atomic size mismatch may play a secondary role. To further show the utility of these rules, we next move to make and then test a prediction for Ni-rich systems. Ni-Zr is particularly promising, as it adheres to all selection criteria (see Table 2). Ni-Zr has demonstrated good glass forming ability and also experiences deep eutectics, similar to the Cu-Zr and Cu-Hf systems [82]. Ni-Zr has a positive $\Delta H^{seg}$ [42], a negative $\Delta H^{mix}$ [83], and an atomic radius mismatch of 29%. A Ni-5.5 at.% Zr alloy was deposited under similar sputtering conditions, with deposition details presented in Table 1. All annealing treatments followed those presented in the Methods section.

Figure 7(a) shows a BF TEM image of the Ni-Zr alloy after the various heat treatments. The inset gives the SAED pattern, with the Ni FCC rings being clearly visible and no other phases detected. Again, despite concerted efforts to induced grain coarsening, the average grain size remained in the nanocrystalline range at 41 nm. The EDS elemental map in Figure 7(b) highlights the segregation of the Zr dopant to the grain boundaries, confirming the positive $\Delta H^{seg}$ of the system. Figure 8(a) shows an HRTEM image of a 3 nm thick AIF in the Ni-Zr alloy, with Figure 8(b) displaying the accompanying EDS line profile scan for this interface. The line profile confirms the elevated Zr concentration in the AIF, reaching a maximum concentration of 21 at.% Zr and dropping back down once inside the grains. While this local percentage is in range for intermetallic formation according to the Ni-Zr phase diagram, no second phases were detected in the SAED pattern. The FFT images confirm the crystalline nature of the two grains and the amorphous nature of the intergranular film. Similar to the Cu-Zr and Cu-Hf systems, the thickness of the AIFs in Ni-Zr were always constant along a given grain boundary, pointing toward thermodynamic equilibrium of the film. Ultimately, the Ni-Zr system matched with the prediction that AIFs will form.

Our simple materials selection rules can also be used to make predictions for a wide variety of alloy systems. Using the $\Delta H^{seg}$ modeling estimations from Murdoch and Schuh [42] in conjunction with $\Delta H^{mix}$ calculated values from Atwater and Darling [63], we present a range of predictions in Table 5 for numerous binary metallic alloy combinations. Blue squares in this table have a positive $\Delta H^{seg}$ and a negative $\Delta H^{mix}$, and are thus predicted to be possible nanoscale AIF formers. Red squares have a positive $\Delta H^{seg}$ and a positive $\Delta H^{mix}$, and are thus predicted to have dopant segregation but only form ordered complexions. Gray squares with an "X" have a negative $\Delta H^{seg}$ and are therefore predicted to have dopant depletion at the interfaces (i.e., dopants prefer to be located inside of the grains). Black squares indicate self-doping (e.g., Al in an Al lattice) or lack of available data to make a prediction. A dot indicates that the alloy has an atomic radius mismatch greater than 12%, as calculated using Equation 2. Other sources were also used to further confirm the enthalpy parameter values where applicable [41, 50, 78, 84-86]. It is worth noting that we do not explicitly treat any competition for dopants from second phase formation here, which can add an additional complication. It is possible that the magnitude of $\Delta H^{mix}$, represented by a relative color scale for both positive and negative values in Table 5, may also be practically important, since very negative values may lead to intermetallic formation that removes dopants from the grain boundaries. Alloys with known ability to form metallic glasses are also particularly promising targets for AIFs and can be used to pinpoint some alloys with great potential. For example, Fe-Ti, Co-Nb, and Ni-Nb have demonstrated good glass forming ability [87, 88] as well as positive $\Delta H^{seg}$ and negative $\Delta H^{mix}$ values, and thus are excellent candidates to form nanoscale AIFs.

It is worth noting that the predictive potential of Table 5 is only as good as the data used to find the thermodynamic parameters. We use the work of Murdoch and Schuh [42] and Atwater and Darling [63] because these are the most complete databases available, but these are not infallible. For example, the $\Delta H^{seg}$ modeling estimations for Cu-Nb and Cu-Mo indicate dopant depletion at the grain boundary, which contradicts the experimental data collected here for these alloys. Thus, any interest in a particular system is best served by the accurate calculation or measurement of the $\Delta H^{seg}$ and $\Delta H^{mix}$ for that exact alloy.

## 4. Conclusions

In this study, a variety of binary Cu-rich alloys and their respective grain boundary structures were evaluated in order to define a relationship between material parameters and the ability to sustain nanoscale AIFs. Four alloys encompassing a range of parameter combinations (Cu-Zr, Cu-Hf, Cu-Nb, and Cu-Mo) were created using sputter deposition and processed to encourage dopant segregation to grain boundaries and grain boundary structure transformation. Analysis of the results from these alloys revealed a pattern of materials selection criteria to predict grain boundary composition and structure. These criteria were then applied to predict and confirm nanoscale AIF formation in Ni-Zr, as well as make predictions for a number of binary transition alloy combinations. The following specific conclusions can be made:

- $\Delta H^{seg}$ and $\Delta H^{mix}$ were found to be the primary determining factors behind the complexion type formed. Other factors that contribute to BMG stability, such as atomic radius mismatch and the usage of three or more elements are secondary at best and require further research to understand their role in AIF formation.

- A positive $\Delta H^{seg}$ coupled with a negative $\Delta H^{mix}$ promotes nanoscale AIF formation in polycrystalline binary metallic alloys. These AIFs were readily observed in Cu-Zr, Cu-Hf, and Ni-Zr.

- A positive $\Delta H^{seg}$ coupled with a positive $\Delta H^{mix}$ promotes ordered grain boundary complexions, where there is also the potential for dopant clustering and phase separation in this scenario. Cu-Nb and Cu-Mo demonstrated doped yet ordered grain boundary structures.

The key conclusion emerging from this study is the development of general materials selection rules for nanoscale AIF formation. Complexion type is determined by the presence of the required dopant element at the grain boundary and the ability of the grain boundary to assume the desired amorphous or crystalline structure. These findings can be leveraged to avoid undesirable material behaviors such as grain boundary embrittlement, and to realize improvements to material behavior such as increased damage tolerance, ductility, and accelerated diffusion.

**Acknowledgments**

This study was supported by the U.S. Department of Energy, Office of Basic Energy Sciences, Materials Science and Engineering Division under Award No. DE-SC0014232. SEM/FIB work was performed at the UC Irvine Materials Research Institute (IMRI), using instrumentation funded in part by the National Science Foundation Center for Chemistry at the Space-Time Limit (CHE-082913). The authors also acknowledge the use of instruments at the Electron Imaging Center for NanoMachines supported by NIH (1S10RR23057 to ZHZ) and CNSI at UCLA.

# References

[1] K. Kumar, H. Van Swygenhoven, S. Suresh, Mechanical behavior of nanocrystalline metals and alloys, Acta Materialia 51(19) (2003) 5743-5774.
[2] Y.S. Zhang, Z. Han, K. Wang, K. Lu, Friction and wear behaviors of nanocrystalline surface layer of pure copper, Wear 260(9) (2006) 942-948.
[3] C. Becquart, C. Domain, Modeling microstructure and irradiation effects, Metallurgical and Materials Transactions A 42(4) (2011) 852-870.
[4] T. Watanabe, Grain boundary engineering: historical perspective and future prospects, Journal of Materials Science 46(12) (2011) 4095-4115.
[5] S.J. Dillon, M.P. Harmer, J. Luo, Grain boundary complexions in ceramics and metals: An overview, JOM 61(12) (2009) 38-44.
[6] S.J. Dillon, M. Tang, W.C. Carter, M.P. Harmer, Complexion: A new concept for kinetic engineering in materials science, Acta Materialia 55(18) (2007) 6208-6218.
[7] X. Shi, J. Luo, Developing grain boundary diagrams as a materials science tool: A case study of nickel-doped molybdenum, Physical Review B 84(1) (2011).
[8] P.R. Cantwell, M. Tang, S.J. Dillon, J. Luo, G.S. Rohrer, M.P. Harmer, Grain boundary complexions, Acta Materialia 62 (2014) 1-48.
[9] M. Tang, W.C. Carter, R.M. Cannon, Grain boundary order-disorder transitions, Journal of Materials Science 41(23) (2006) 7691-7695.
[10] A. Kundu, K.M. Asl, J. Luo, M.P. Harmer, Identification of a bilayer grain boundary complexion in Bi-doped Cu, Scripta Materialia 68(2) (2013) 146-149.
[11] J. Luo, H. Cheng, K.M. Asl, C.J. Kiely, M.P. Harmer, The role of a bilayer interfacial phase on liquid metal embrittlement, Science 333(6050) (2011) 1730-1733.
[12] W. Sigle, G. Richter, M. Rühle, S. Schmidt, Insight into the atomic-scale mechanism of liquid metal embrittlement, Applied Physics Letters 89(12) (2006) 121911.
[13] Z. Pan, T.J. Rupert, Amorphous intergranular films as toughening structural features, Acta Materialia 89 (2015) 205-214.
[14] Z. Pan, T.J. Rupert, Damage nucleation from repeated dislocation absorption at a grain boundary, Computational Materials Science 93 (2014) 206-209.
[15] C. Brandl, T.C. Germann, A. Misra, Structure and shear deformation of metallic crystalline–amorphous interfaces, Acta Materialia 61(10) (2013) 3600-3611.
[16] J.E. Ludy, T.J. Rupert, Amorphous intergranular films act as ultra-efficient point defect sinks during collision cascades, Scripta Materialia 110 (2016) 37-40.
[17] A. Khalajhedayati, Z. Pan, T.J. Rupert, Manipulating the interfacial structure of nanomaterials to achieve a unique combination of strength and ductility, Nature Communications 7 (2016).
[18] A. Khalajhedayati, T.J. Rupert, High-temperature stability and grain boundary complexion formation in a nanocrystalline Cu-Zr alloy, JOM 67(12) (2015) 2788-2801.
[19] J. Luo, Liquid-like interface complexion: From activated sintering to grain boundary diagrams, Current Opinion in Solid State and Materials Science 12(5-6) (2008) 81-88.
[20] Y. Wang, J. Li, A.V. Hamza, T.W. Barbee Jr., Ductile crystalline-amorphous nanolaminates, Proceedings of the National Academy of Sciences of the United States of America 104(27) (2007) 11155-60.
[21] M. Faraday, Note on regelation, Proceedings of the Royal Society of London 10 (1859) 440-450.

[22] E.W. Hart, Two-dimensional phase transformation in grain boundaries, Scripta Metallurgica 2(3) (1968) 179-182.
[23] J. Cahn, Transitions and phase equilibria among grain boundary structures, Le Journal de Physique Colloques 43(C6) (1982) C6-199-C6-213.
[24] A.P. Sutton, R.W. Balluffi, H. Lüth, J.M. Gibson, Interfaces in crystalline materials and surfaces and interfaces of solid materials, Physics Today 49 (1996) 88.
[25] S.J. Dillon, M.P. Harmer, Multiple grain boundary transitions in ceramics: a case study of alumina, Acta Materialia 55(15) (2007) 5247-5254.
[26] S. Divinski, M. Lohmann, C. Herzig, B. Straumal, B. Baretzky, W. Gust, Grain-boundary melting phase transition in the Cu− Bi system, Physical Review B 71(10) (2005) 104104.
[27] M. Tang, W.C. Carter, R.M. Cannon, Grain boundary transitions in binary alloys, Phys. Rev. Lett. 97(7) (2006) 075502.
[28] D.R. Clarke, On the detection of thin intergranular films by electron microscopy, Ultramicroscopy 4(1) (1979) 33-44.
[29] J. Luo, X. Shi, Grain boundary disordering in binary alloys, Applied Physics Letters 92(10) (2008) 101901.
[30] W.H. Zachariasen, The atomic arrangement in glass, Journal of the American Chemical Society 54(10) (1932) 3841-3851.
[31] A.R. Cooper, W.H. Zachariasen - the melody lingers on, Journal of Non-Crystalline Solids 49(1) (1982) 1-17.
[32] A. Inoue, Stabilization of metallic supercooled liquid and bulk amorphous alloys, Acta Materialia 48(1) (2000) 279-306.
[33] A. Ziegler, J.C. Idrobo, M.K. Cinibulk, C. Kisielowski, N.D. Browning, R.O. Ritchie, Interface structure and atomic bonding characteristics in silicon nitride ceramics, Science 306(5702) (2004) 1768-1770.
[34] M.K. Cinibulk, H.J. Kleebe, Effects of oxidation on intergranular phases in silicon nitride ceramics, Journal of Materials Science 28(21) (1993) 5775-5782.
[35] D.B. Williams, C.B. Carter, Transmission Electron Microscopy: A Textbook for Materials Science, Springer, New York, 1996.
[36] C. Koch, I. Ovid'ko, S. Seal, S. Veprek, Structural Nanocrystalline Materials: Fundamentals and Applications Cambridge University Press, New York, 2007.
[37] J.A. Thornton, Influence of substrate temperature and deposition rate on structure of thick sputtered Cu coatings, Journal of Vacuum Science and Technology 12(4) (1975) 830-835.
[38] C. Thompson, Structure evolution during processing of polycrystalline films, Annual Review of Materials Science 30(1) (2000) 159-190.
[39] Q. Jin, D. Wilkinson, G. Weatherly, Determination of grain-boundary film thickness by the fresnel fringe imaging technique, Journal of the European Ceramic Society 18(15) (1998) 2281-2286.
[40] T. Hsieh, R. Balluffi, Experimental study of grain boundary melting in aluminum, Acta Metallurgica 37(6) (1989) 1637-1644.
[41] C.W. Corti, Sintering aids in powder metallurgy, Platinum Metals Review 30(4) (1986) 184-195.
[42] H.A. Murdoch, C.A. Schuh, Estimation of grain boundary segregation enthalpy and its role in stable nanocrystalline alloy design, Journal of Materials Research 28(16) (2013) 2154-2163.
[43] J. Weissmüller, Alloy effects in nanostructures, Nanostructured Materials 3(1-6) (1993) 261-272.

[44] C.C. Koch, R.O. Scattergood, K.A. Darling, J.E. Semones, Stabilization of nanocrystalline grain sizes by solute additions, Journal of Materials Science 43(23-24) (2008) 7264-7272.
[45] A.R. Kalidindi, T. Chookajorn, C.A. Schuh, Nanocrystalline materials at equilibrium: A thermodynamic review, JOM 67(12) (2015) 2834-2843.
[46] T.J. Rupert, The role of complexions in metallic nano-grain stability and deformation, Current Opinion in Solid State and Materials Science 20(5) (2016) 257-267.
[47] J.R. Trelewicz, C.A. Schuh, Grain boundary segregation and thermodynamically stable binary nanocrystalline alloys, Physical Review B 79(9) (2009) 094112.
[48] H.A. Murdoch, C.A. Schuh, Stability of binary nanocrystalline alloys against grain growth and phase separation, Acta Materialia 61(6) (2013) 2121-2132.
[49] T. Chookajorn, C.A. Schuh, Thermodynamics of stable nanocrystalline alloys: A Monte Carlo analysis, Physical Review B 89(6) (2014) 064102.
[50] A.R. Kalidindi, C.A. Schuh, Stability criteria for nanocrystalline alloys, Acta Materialia 132 (2017) 128-137.
[51] K.A. Darling, M.A. Tschopp, B.K. VanLeeuwen, M.A. Atwater, Z.K. Liu, Mitigating grain growth in binary nanocrystalline alloys through solute selection based on thermodynamic stability maps, Computational Materials Science 84 (2014) 255-266.
[52] Z. Pan, T.J. Rupert, Spatial variation of short-range order in amorphous intergranular complexions, Computational Materials Science 131 (2017) 62-68.
[53] W.H. Wang, C. Dong, C.H. Shek, Bulk metallic glasses, Materials Science and Engineering: R: Reports 44(2-3) (2004) 45-89.
[54] D. Wang, Y. Li, B. Sun, M. Sui, K. Lu, E. Ma, Bulk metallic glass formation in the binary Cu–Zr system, Applied Physics Letters 84(20) (2004) 4029-4031.
[55] W. Xing, A.R. Kalidindi, C.A. Schuh, Preferred nanocrystalline configurations in ternary and multicomponent alloys, Scripta Materialia 127 (2017) 136-140.
[56] N.N. Greenwood, A. Earnshaw, Chemistry of the Elements (Second Edition), Butterworth-Heinemann, Oxford, 1997.
[57] K. Zhang, B. Dice, Y. Liu, J. Schroers, M.D. Shattuck, C.S. O'Hern, On the origin of multi-component bulk metallic glasses: Atomic size mismatches and de-mixing, The Journal of Chemical Physics 143(5) (2015) 054501.
[58] D.A. Porter, K.E. Easterling, M. Sherif, Phase Transformations in Metals and Alloys, (Revised Reprint), CRC press, New York, 2009.
[59] M. Kapoor, T. Kaub, K.A. Darling, B.L. Boyce, G.B. Thompson, An atom probe study on Nb solute partitioning and nanocrystalline grain stabilization in mechanically alloyed Cu-Nb, Acta Materialia 126 (2017) 564-575.
[60] Z. Pan, T.J. Rupert, Formation of ordered and disordered interfacial films in immiscible metal alloys, Scripta Materialia 130 (2017) 91-95.
[61] N.Q. Vo, J. Schäfer, R.S. Averback, K. Albe, Y. Ashkenazy, P. Bellon, Reaching theoretical strengths in nanocrystalline Cu by grain boundary doping, Scripta Materialia 65(8) (2011) 660-663.
[62] N.Q. Vo, S.W. Chee, D. Schwen, X. Zhang, P. Bellon, R.S. Averback, Microstructural stability of nanostructured Cu alloys during high-temperature irradiation, Scripta Materialia 63(9) (2010) 929-932.
[63] M.A. Atwater, K.A. Darling, A Visual Library of Stability in Binary Metallic Systems: The Stabilization of Nanocrystalline Grain Size by Solute Addition: Part 1, Army Research Lab Aberdeen Proving Ground MD Weapons and Materials Research Directorate, 2012.

[64] O. Kleppa, S. Watanabe, Thermochemistry of alloys of transition metals: Part III. Copper-silver,-titanium,-zirconium, and-hafnium at 1373 K, Journal of Electronic Materials 20(12) (1991) 391-401.
[65] L. Zhang, E. Martinez, A. Caro, X.Y. Liu, M.J. Demkowicz, Liquid-phase thermodynamics and structures in the Cu–Nb binary system, Modelling and Simulation in Materials Science and Engineering 21(2) (2013) 025005.
[66] C.J. Smithells, W.F. Gale, T.C. Totemeier, Smithells Metals Reference Book (8th Edition), Elsevier Butterworth-Heinemann, Boston, 2004.
[67] D.H. Kang, H. Zhang, H. Yoo, H.H. Lee, S. Lee, G.W. Lee, H. Lou, X. Wang, Q. Cao, D. Zhang, J. Jiang, Interfacial free energy controlling glass-forming ability of Cu-Zr alloys, Sci. Rep. 4 (2014) 5167.
[68] A. Inoue, W. Zhang, Formation, thermal stability and mechanical properties of Cu-Zr and Cu-Hf binary glassy alloy rods, Materials Transactions 45(2) (2004) 584-587.
[69] M. Klinger, A. Jager, Crystallographic Tool Box (CrysTBox): Automated tools for transmission electron microscopists and crystallographers, J. Appl. Crystallogr. 48(Pt 6) (2015) 2012-2018.
[70] C. Gammer, C. Mangler, C. Rentenberger, H.P. Karnthaler, Quantitative local profile analysis of nanomaterials by electron diffraction, Scripta Materialia 63(3) (2010) 312-315.
[71] Z. Pan, T.J. Rupert, Effect of grain boundary character on segregation-induced structural transitions, Physical Review B 93(13) (2016) 134113.
[72] K.A. Darling, B.K. VanLeeuwen, C.C. Koch, R.O. Scattergood, Thermal stability of nanocrystalline Fe–Zr alloys, Materials Science and Engineering: A 527(15) (2010) 3572-3580.
[73] A.J. Detor, C.A. Schuh, Microstructural evolution during the heat treatment of nanocrystalline alloys, Journal of Materials Research 22(11) (2007) 3233-3248.
[74] M.N. Polyakov, T. Chookajorn, M. Mecklenburg, C.A. Schuh, A.M. Hodge, Sputtered Hf–Ti nanostructures: A segregation and high-temperature stability study, Acta Materialia 108 (2016) 8-16.
[75] T. Chookajorn, H.A. Murdoch, C.A. Schuh, Design of stable nanocrystalline alloys, Science 337(6097) (2012) 951-954.
[76] F. Abdeljawad, P. Lu, N. Argibay, B.G. Clark, B.L. Boyce, S.M. Foiles, Grain boundary segregation in immiscible nanocrystalline alloys, Acta Materialia 126 (2017) 528-539.
[77] T. Frolov, Y. Mishin, Stable nanocolloidal structures in metallic systems, Phys. Rev. Lett. 104(5) (2010) 055701.
[78] K.A. Darling, A.J. Roberts, Y. Mishin, S.N. Mathaudhu, L.J. Kecskes, Grain size stabilization of nanocrystalline copper at high temperatures by alloying with tantalum, Journal of Alloys and Compounds 573 (2013) 142-150.
[79] E. Botcharova, J. Freudenberger, L. Schultz, Mechanical and electrical properties of mechanically alloyed nanocrystalline Cu–Nb alloys, Acta Materialia 54(12) (2006) 3333-3341.
[80] T. Frolov, K.A. Darling, L.J. Kecskes, Y. Mishin, Stabilization and strengthening of nanocrystalline copper by alloying with tantalum, Acta Materialia 60(5) (2012) 2158-2168.
[81] T. Gladman, On the theory of the effect of precipitate particles on grain growth in metals, Proceedings of the Royal Society of London A: Mathematical, Physical and Engineering Sciences 294(1438) (1966) 298-309.
[82] Y.D. Dong, G. Gregan, M.G. Scott, Formation and stability of nickel-zirconium glasses, Journal of Non-Crystalline Solids 43(3) (1981) 403-415.

[83] M. Rösner-Kuhn, J. Qin, K. Schaefers, U. Thiedemann, M.G. Frohberg, Temperature dependence of the mixing enthalpy and excess heat capacity in the liquid system nickel-zirconium, International Journal of Thermophysics 17(4) (1996) 959–966.

[84] G. Ghosh, C. Kantner, G.B. Olson, Thermodynamic modeling of the Pd-X (X=Ag, Co, Fe, Ni) systems, Journal of Phase Equilibria 20(3) (1999) 295.

[85] Y.M. Kim, Y.H. Shin, B.J. Lee, Modified embedded-atom method interatomic potentials for pure Mn and the Fe–Mn system, Acta Materialia 57(2) (2009) 474-482.

[86] O.J. Kleppa, L. Topor, Thermochemistry of binary liquid gold alloys: The systems (Au + Cr), (Au + V), (Au + Ti), and (Au + Sc) at 1379 K, Metallurgical Transactions A 16(1) (1985) 93-99.

[87] C.J. Lin, F. Spaepen, Nickel-niobium alloys obtained by picosecond pulsed laser quenching, Acta Metallurgica 34(7) (1986) 1367-1375.

[88] F. Spaepen, Interdiffusion in amorphous metallic artificial multilayers, Materials Science and Engineering 97 (1988) 403-408.

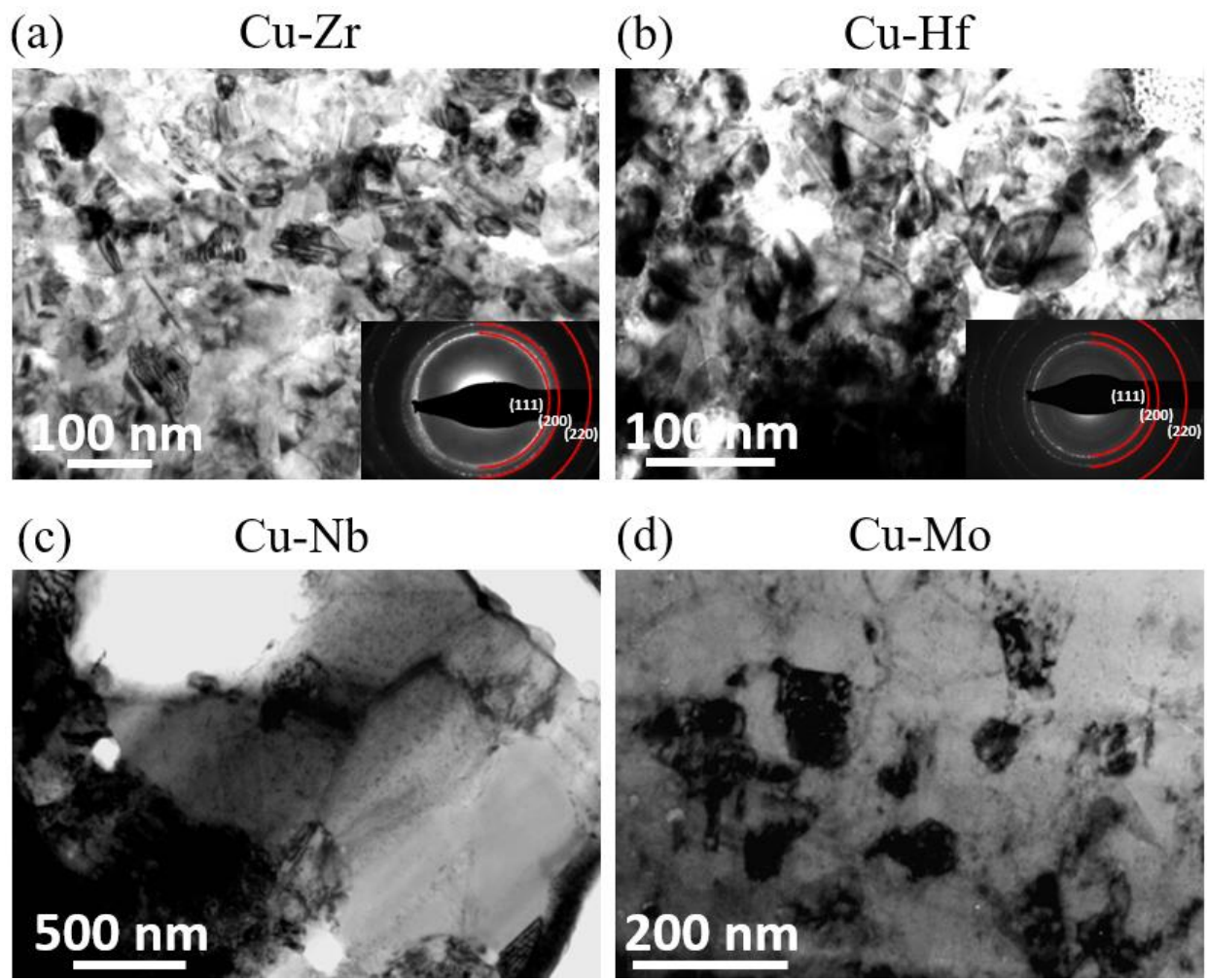

**Figure 1.** Bright field TEM images of the (a) Cu-Zr, (b) Cu-Hf, (c) Cu-Nb and (d) Cu-Mo films after completion of all heat treatment steps. The insets show the SAED pattern for (a) Cu-Zr and (b) Cu-Hf, where only single phase FCC diffraction rings are observed.

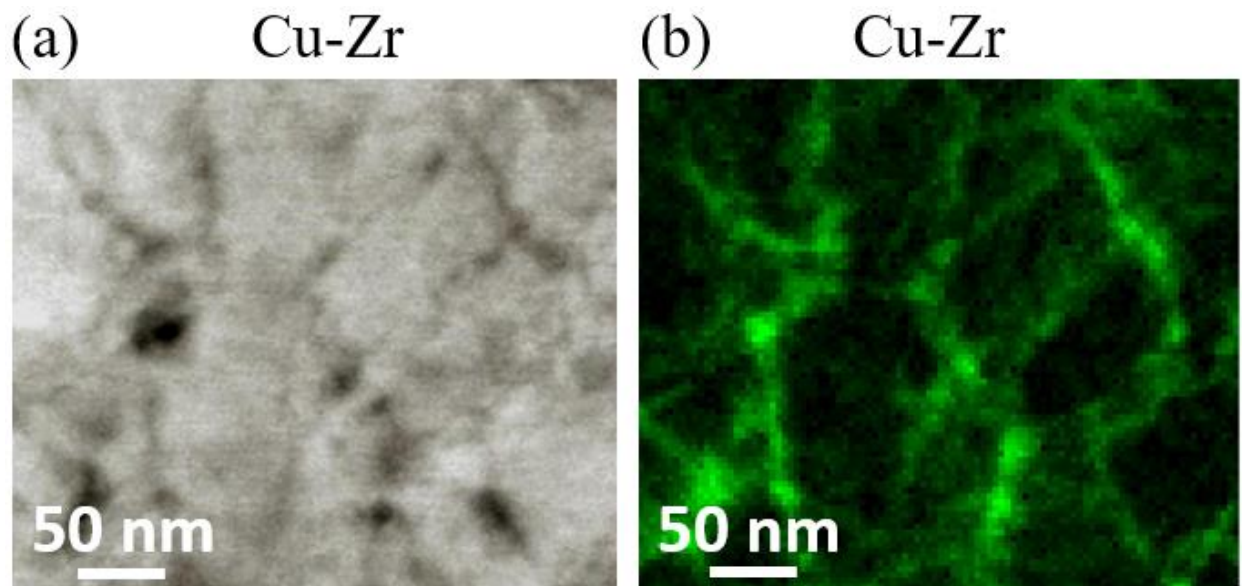

**Figure 2.** (a) STEM image and (b) the corresponding EDS mapping of the (b) Cu-Zr (green) alloy, after completion of all heat treatment steps. Green regions correspond to high Zr content, which is present at the grain boundaries.

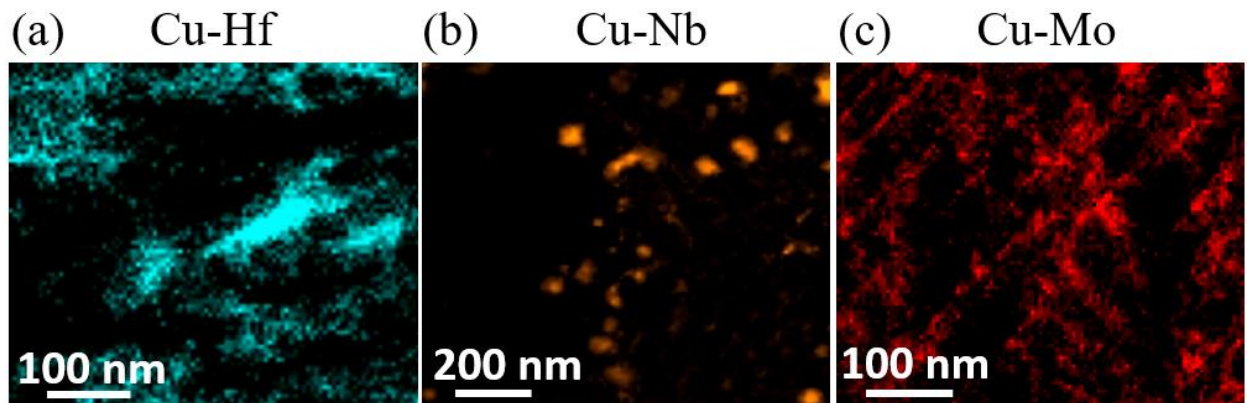

**Figure 3.** EDS mapping of the (a) Cu-Hf (blue), (b) Cu-Nb (orange) and (c) Cu-Mo (red) alloys, after completion of all heat treatment steps. Blue, orange, and red regions correspond to high levels of Hf, Nb, and Mo, respectively.

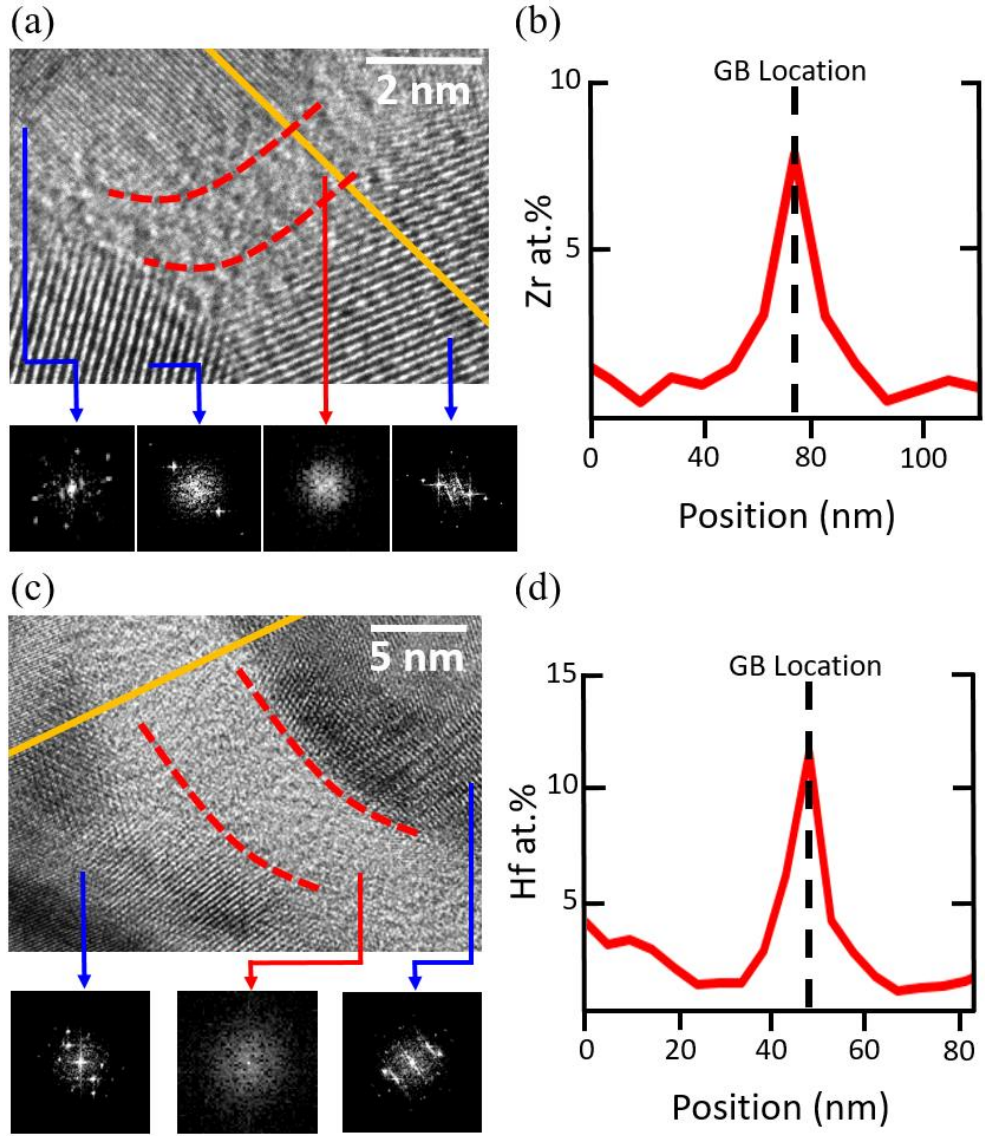

**Figure 4.** High resolution TEM images of amorphous intergranular films in the (a) Cu-Zr and (c) Cu-Hf samples, with FFT images shown in the insets. EDS line profile scans across (b) the Cu-Zr sample and (d) the Cu-Hf samples are also shown. The yellow lines in (a) and (c) give the scan locations, with the grain boundary (GB) location marked on the line profiles in (b) and (d).

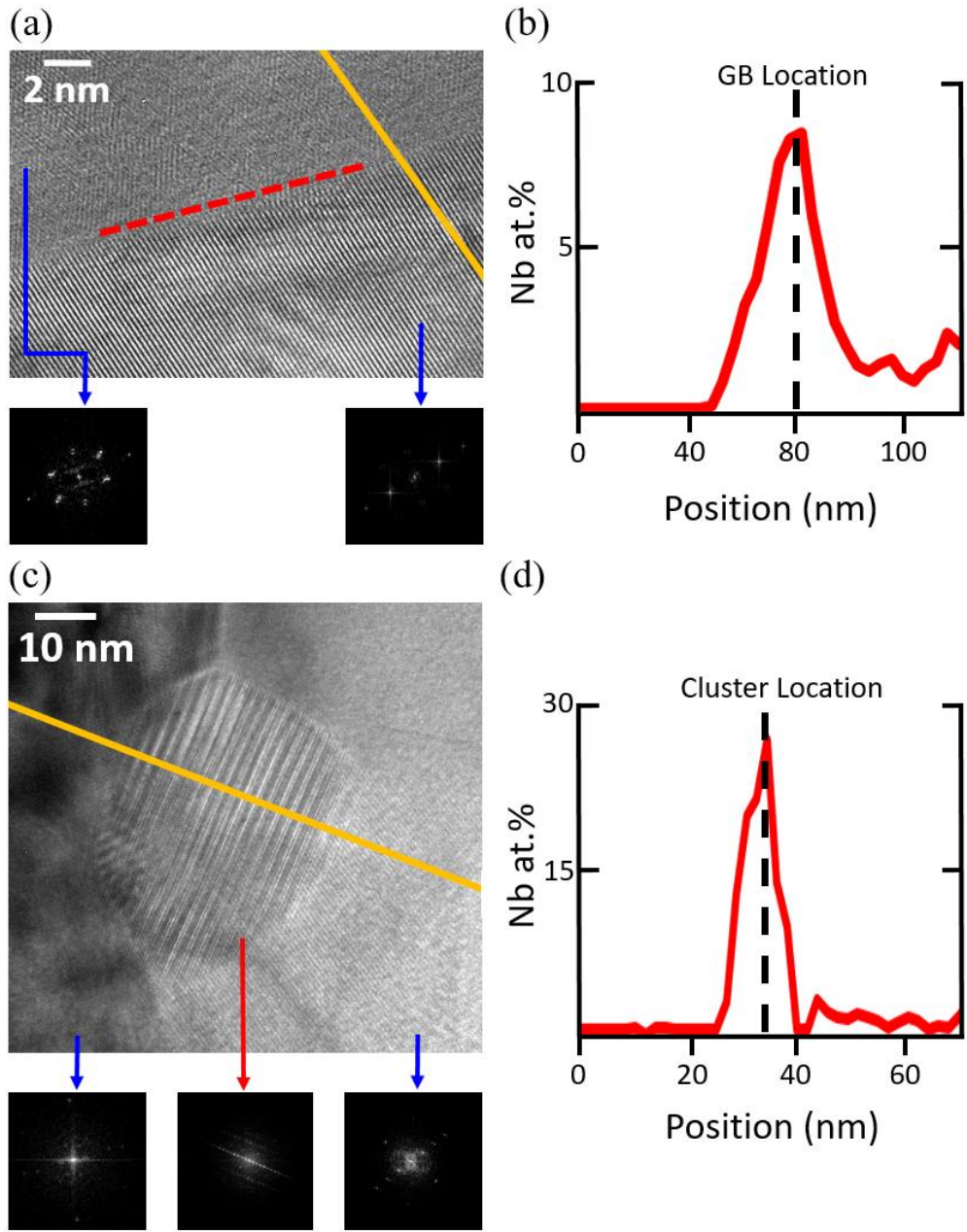

**Figure 5.** HRTEM images from Cu-Nb of (a) an ordered grain boundary and (c) a Nb-rich cluster located along a grain boundary. FFT images shown in the insets are sampled across the grain boundary film and the Nb-rich cluster, as well as the grain interiors. EDS line profile scans are given across the (b) grain boundary and (d) Nb-rich cluster. The yellow lines in (a) and (c) give the scan locations, with the grain boundary (GB) and cluster location marked on the line profiles in (b) and (d) respectively.

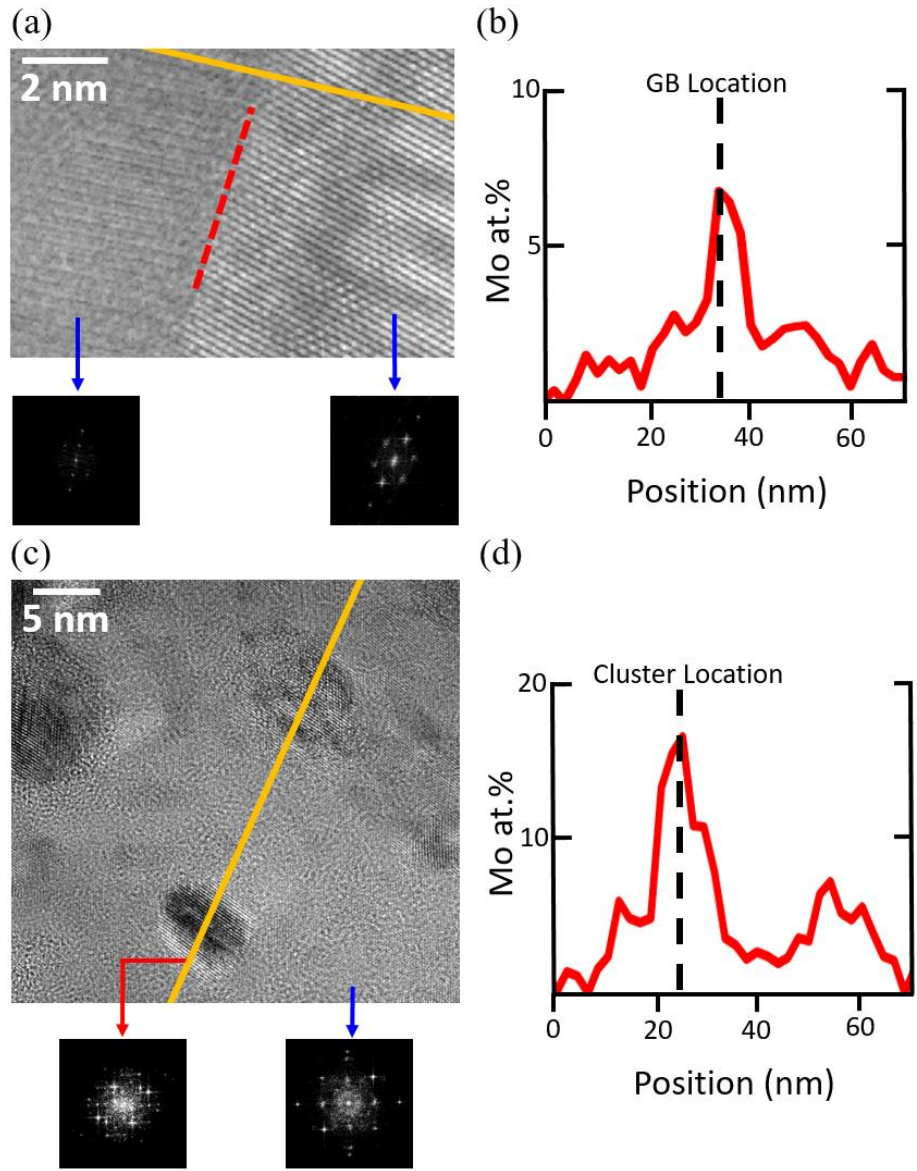

**Figure 6.** HRTEM images from Cu-Mo of (a) an ordered grain boundary and (c) a Mo-rich cluster. FFT insets are sampled across the grain boundary and Mo-rich cluster. EDS line profile scans are given across the (b) grain boundary and (d) Mo-rich cluster. The yellow lines in (a) and (c) give the scan locations, with the grain boundary (GB) and cluster location marked on the line profiles in (b) and (d) respectively.

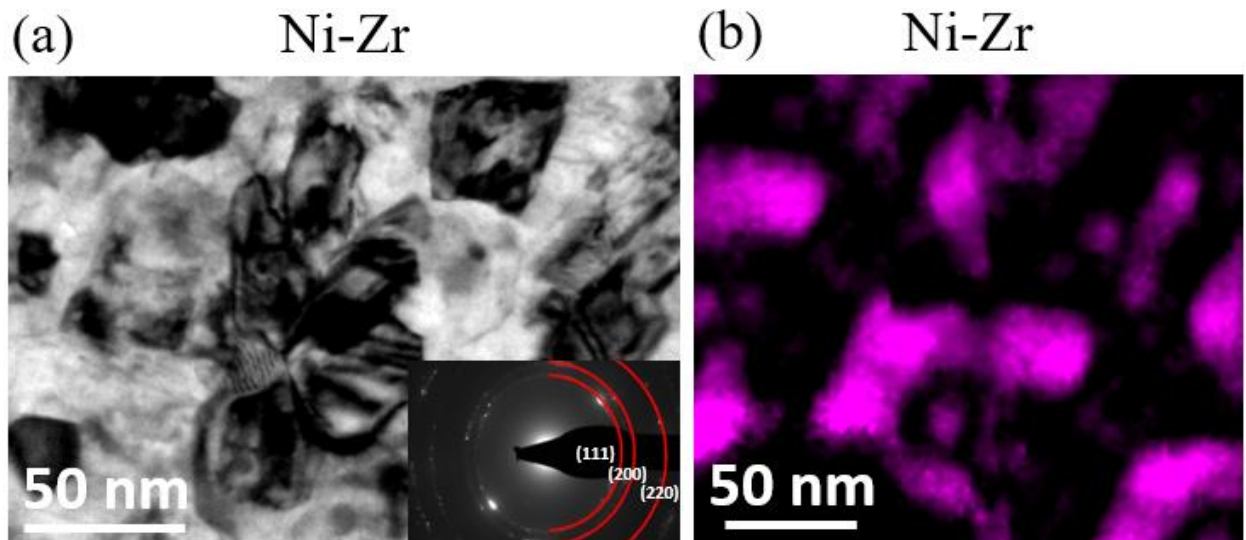

**Figure 7.** (a) Bright field TEM image of Ni-Zr after completion of all heat treatment steps, with an SAED inset. (b) Associated EDS mapping of the heat treated Ni-Zr with Zr denoted by pink.

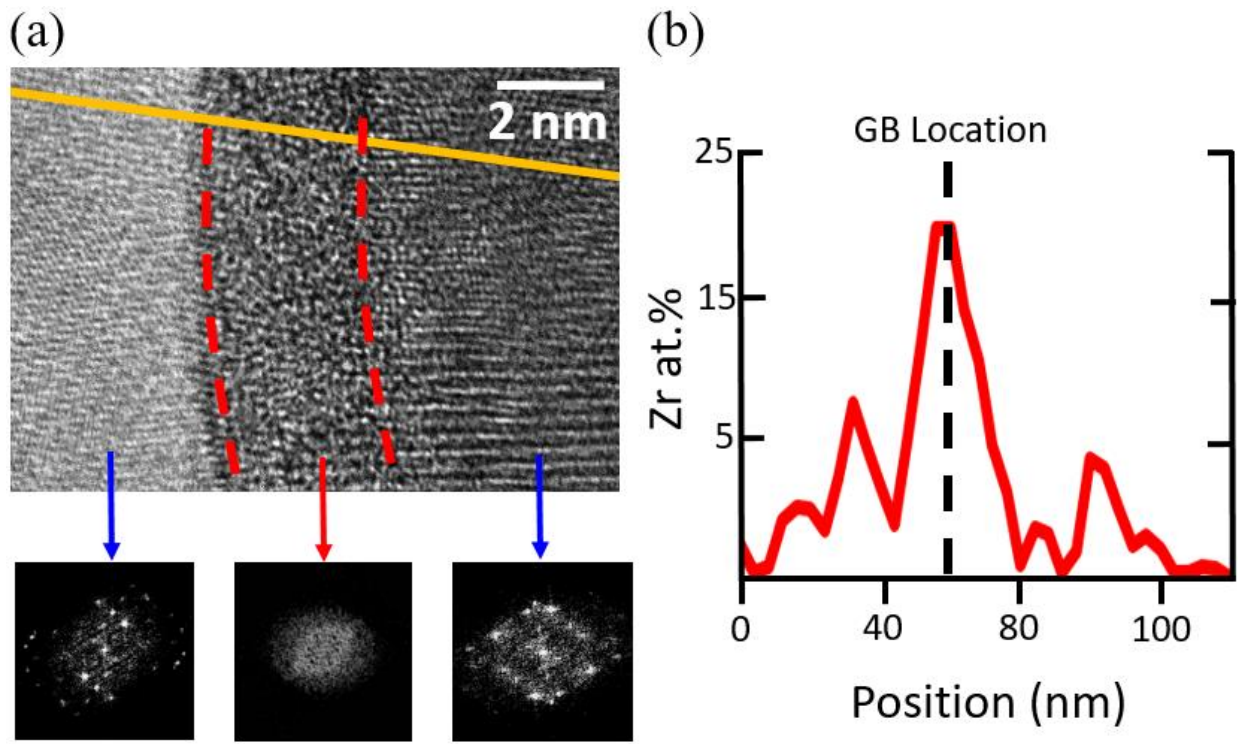

**Figure 8.** High resolution TEM image of (a) an amorphous intergranular film in Ni-Zr with FFT insets sampled across the boundary structure. The EDS line profile scan across the amorphous intergranular film is shown (b). The yellow line in (a) gives the scan location, with the grain boundary (GB) location marked on the line profile in (b).

| Alloy | Substrate | Dep. Temp. (°C) | Solute/ Solvent Dep. Power (W) | Ar base Pressure (mTorr) | Dep. Rate (Å/sec) | Avg. Film Thickness (μm) | Post Quench (at. %) | Avg. Grain Size (nm) | $0.92T_{solidus}$ (°C) |
|---|---|---|---|---|---|---|---|---|---|
| Cu-Zr | Cu | 400 | 75/150 | 1.5 | 1.8 | 1.94 | 4.3 | 99 ± 29 | 900 |
| Cu-Hf | Cu | 400 | 75/150 | 1.5 | 1.7 | 2.04 | 6.2 | 47 ± 12 | 915 |
| Cu-Nb | Cu | 400 | 75/150 | 1.5 | 1.6 | 1.92 | 2.7 | 468 ± 185 | 1000 |
| Cu-Mo | Cu | 400 | 75/150 | 1.5 | 1.6 | 1.82 | 3.3 | 85 ± 26 | 1000 |
| Ni-Zr | Ni | 400 | 75/150 | 1.5 | 0.9 | 1.34 | 5.5 | 40 ± 12 | 1100 |

**Table 1.** The sputter deposition parameters including substrate, deposition temperature, power, base pressure and deposition rate. Also included is the resultant film thickness, dopant percentage and final grain size of each alloy after all thermal processing treatments were completed.

| Alloy | $\Delta H^{seg}$ | $\Delta H^{mix}$ | Atomic Radius Mismatch (%) | Grain Boundary Prediction |
|---|---|---|---|---|
| Cu-Zr | Positive | Negative | 25 | AIF |
| Cu-Hf | Positive | Negative | 24 | AIF |
| Cu-Nb | Positive | Positive | 14 | Ordered |
| Cu-Mo | Positive | Positive | 9 | Ordered |
| Ni-Zr | Positive | Negative | 29 | AIF |

**Table 2.** The thermodynamic variables and predictions for complexion type for the binary metallic alloys. Alloys with a positive $\Delta H^{seg}$ coupled with a negative $\Delta H^{mix}$ are predicted to have AIF formation. In contrast, those alloys having a positive $\Delta H^{seg}$ coupled with a positive $\Delta H^{mix}$ are predicted to have ordered grain boundaries. An atomic radius mismatch >12% promotes BMG formation and is also evaluated for its influence on grain boundary structure.

| Alloy | Dopant Segregation to Grain Boundary? | Complexion Type Found |
|---|---|---|
| Cu-Zr | Yes | AIF |
| Cu-Hf | Yes | AIF |
| Cu-Nb | Yes | Ordered |
| Cu-Mo | Yes | Ordered |
| Ni-Zr | Yes | AIF |

**Table 3:** A summary of the final results for both the Cu-rich and Ni-rich systems. All of the systems experienced dopant segregation. Cu-Zr and Cu-Hf both had AIF formation, while Cu-Nb and Cu-Mo had ordered grain boundaries. Using this knowledge, Ni-Zr was predicted to contain AIFs, which was confirmed.

| Primary Element | Dopant | $\Delta H^{seg}$ | $\Delta H^{mix}$ | Atomic Radius Mismatch (%) | Observed Behavior | Complexion Structure |
|---|---|---|---|---|---|---|
| Ni | Bi | + | + | * | GB embrittlement [11] | Ordered |
| Cu | Bi | + | + | * | GB embrittlement [10] | Ordered |
| Al | Ga | + | + | * | GB embrittlement [12] | Ordered |
| Mo | Fe | + | − | 10 | Activated sintering [41] | AIF |
| Mo | Co | + | − | 11 | Activated sintering [41] | AIF |
| Mo | Ni | + | − | 12 | Activated sintering [41] | AIF |
| Mo | Rh | + | − | 4 | Activated sintering [41] | AIF |
| Mo | Pd | + | − | 2 | Activated sintering [41] | AIF |
| Mo | Pt | + | − | <1 | Activated sintering [41] | AIF |
| W | Co | + | − | 11 | Activated sintering [29] | AIF |
| W | Ni | + | − | 12 | Activated sintering [29, 41] | AIF |
| W | Ru | + | − | 4 | Activated sintering [41] | AIF |
| W | Rh | + | − | 4 | Activated sintering [41] | AIF |
| W | Pd | + | − | 2 | Activated sintering [29] | AIF |
| W | Pt | + | − | <1 | Activated sintering [41] | AIF |
| W | Cu | + | + | 9 | No activated sintering [29] | Ordered |

**Table 4**. Additional binary alloys that have exhibited behavior that can be potentially attributed to complexion formation. All of the alloys have a positive $\Delta H^{seg}$, meaning dopant segregation to the grain boundary is energetically favorable. Ni-Bi, Cu-Bi and Al-Ga have a positive $\Delta H^{mix}$, which predicts an ordered grain boundary structure (confirmed experimentally) and has been attributed to boundary embrittlement. The Mo and W alloys (except W-Cu) have negative $\Delta H^{mix}$ and experience solid-state activated sintering, behavior which has been attributed to AIFs. In contrast, activated sintering has not been observed for W-Cu, which aligns with the positive $\Delta H^{mix}$ and ordered grain boundaries predicted for these systems. The atomic radius mismatch values calculated using the metallic bonding radii are also under 12% for many of the alloys that experience activated sintering and have AIFs, providing further confirmation that this parameter plays a secondary role in encouraging AIF formation. Those alloys with a (*) have potentially directional bonding which may influence the atomic radius mismatch calculation.

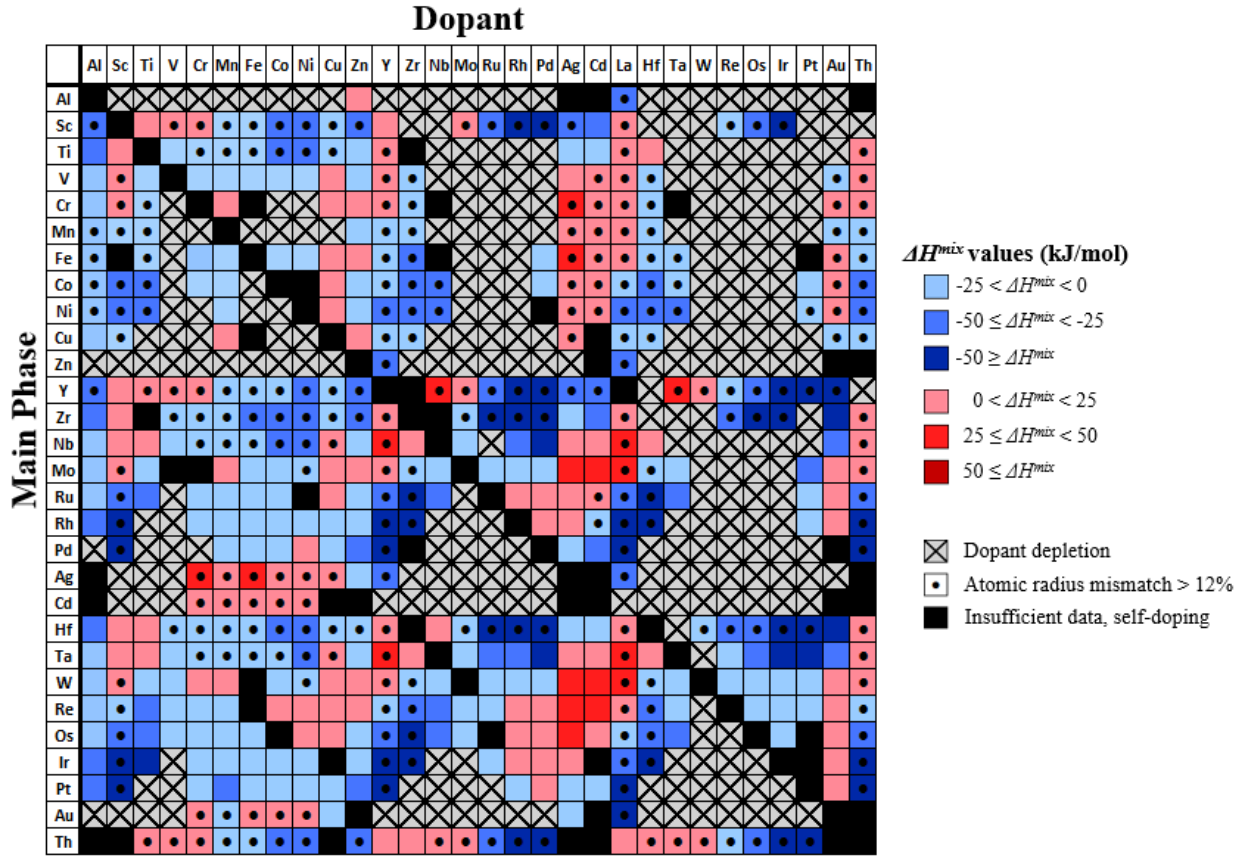

**Table 5**. Binary transition metal alloys evaluated for nanoscale AIF formation. Blue squares denote a positive $\Delta H^{seg}$ and a negative $\Delta H^{mix}$, and are thus predicted to be possible AIF formers. Red squares have a positive $\Delta H^{seg}$ and a positive $\Delta H^{mix}$, and are thus predicted to have dopant segregation and ordered complexions. Gray squares with an "X" have a negative $\Delta H^{seg}$ and are predicted to have dopant depletion at the grain boundary. Black squares indicate self-doping or lack of available data to make a prediction. A dot indicates that the alloy has an atomic radius mismatch greater than 12%. The modeling calculation values for $\Delta H^{seg}$ are gathered from Murdoch and Schuh [42], while $\Delta H^{mix}$ values are gathered from Atwater and Darling [63].